\documentstyle[prl,aps,multicol,epsf]{revtex}

\draft

\begin{document}
\preprint{ }
\title{Structure of Flux Line Lattices with Weak Disorder at Large Length Scales}

\author{Philip Kim, Zhen Yao\cite{zhen}, Cristian A. Bolle \cite{cris}
and Charles M. Lieber}

\address{Division of Engineering and Applied Sciences, Harvard
University, Cambridge, MA 02138}

\maketitle

\begin{abstract}
Dislocation-free decoration images containing up to 80,000
vortices have been obtained on high quality
Bi$_{2}$Sr$_{2}$CaCu$_{2}$O$_{8+x}$ superconducting single
crystals.  The observed flux line lattices are in the random
manifold regime with a roughening exponent of 0.44 for length
scales up to 80-100 lattice constants.  At larger length scales,
the data exhibit nonequilibrium features that persist for
different cooling rates and field histories.

\end{abstract}
\pacs{PACS numbers: 74.60.Ge, 75.10.Nr, 74.60.Ec, 74.72.Hs}

\begin{multicols}{2}

\narrowtext


Recent studies of high temperature superconductors have shown
richness in the phase diagram due to the presence of weak quenched
disorder \cite{[1]}.  Larkin first showed that arbitrarily weak
disorder destroys the long range translational order of flux lines
(FLs) in a lattice \cite{[2]}.  It was recently pointed out that
the Larkin model, which is based on a small displacement expansion
of the disorder potential, cannot be applied to length scales
larger than the correlated volume of the impurity potential termed
the Larkin regime \cite{[3],[4],[5]}.  Beyond the Larkin regime,
the behavior of FLs in the absence of dislocations has been
considered using elastic models \cite{[3],[4],[5]}.  First, FLs
start to behave collectively as an elastic manifold in a random
potential with many metastable states (the random manifold regime)
\cite{[3]}.  In this random manifold regime, the translational
order decreases as a stretched exponential, whereas there is a
more rapid exponential decay in the Larkin regime. At even larger
length scales, when the displacement correlation of FLs become
comparable to the lattice spacing, the random manifold regime
transits to a quasiordered regime where the translational order
decays as a power law \cite{[4],[5]}.

Experimentally, neutron diffraction \cite{[6]} and local Hall
probe measurements \cite{[7]} have shown the existence of an
order-disorder phase transition with increased field, although the
microscopic details of these phases are not clear.  Theoretical
progress describing FLs in the presence of weak disorder has been
made within elastic theory, which proposes the absence of
dislocations at equilibrium \cite{[5],[8],[9]}.  To date, however,
there has been no experimental work addressing the structure of
dislocation-free FL lattices at large length scales.  Previous
magnetic decoration studies \cite{[10],[11]} showed that the
dislocation density decreases and the translational order
increases with increasing magnetic field.  However, only
relatively short-range translational order could be probed in the
previous work due to the finite image size and relatively low
applied fields.

In this paper, we report the first large length scale structural
studies of FLs with measurements extending up to $\sim 300$
lattice constants and fields up to 120 G. Real-space images show
dislocation free regions containing up to the order of $10^{5}$
FLs.  A very low density of dislocations was also observed,
although detailed analysis suggests that the dislocations are not
equilibrium features.  The translational correlation function and
displacement correlator have been calculated from dislocation free
data to examine quantitatively the decay of order.  These results
show a stretched exponential decay of the translational order
indicating that FLs are in the random manifold regime.  The
experimentally determined roughening exponent in the random
manifold regime agrees well with theoretical predictions.

High quality single crystals of
Bi$_{2}$Sr$_{2}$CaCu$_{2}$O$_{8+x}$ (BSCCO) were grown as
described elsewhere \cite{[12]}.  Typically, crystals of $\sim 1
mm\times 1 mm\times 20\mu$m size were mounted on a copper
cold-finger and decorated with thermally evaporated iron clusters
at 4 K. The samples were cooled down to 4 K using different
thermal cycles to test nonequilibrium effects and to achieve as
close an equilibrium configuration of FLs as possible within the
experimental time scale.  The FL structure was imaged after
decoration using a scanning electron microscope equipped with a
4096 x 4096 pixel, 8-bit gray-scale image acquisition system.
Nonlinearity in the system was eliminated using grating standards.
This high-resolution system enabled us to acquire images
containing nearly $10^{5}$ FLs, while maintaining a similar
resolution ($\sim14$ pixels between vortices) to previous studies
of $10^{3}$ FLs. In addition, an iterative Voronoi construction
\cite{[13]} was used to reduce the positioning inaccuracy to 3 \%
of a lattice constant.

Samples were decorated in fields of 70, 80 and 120 G parallel to
the c axis of BSCCO single crystals.  In contrast to the previous
decoration experiments at lower fields \cite{[10],[11]}, we find
that the dislocations are rare at these fields.  The density of
dislocations was $1.7\times 10^{-5}$, $1.4\times 10^{-5}$ and
$3.1\times 10^{-5}$ for 70, 80 and 120 G, respectively, where the
total number of vortices is $\sim 240,000$ for each field. It is
thus trivial to find many large $100\times 100$ $\mu$m$^{2}$
dislocation free regions in the decorated samples. The size of the
largest dislocation free image, which was obtained in a field of
70 G, is $152\times 152$ $\mu$m$^{2}$ with 78,363 vortices
\cite{[14]}. Although a small number of dislocations are detected
in our FL images, this does not imply that they are energetically
favorable at equilibrium. On the contrary, we believe that the
large dislocation-free areas observed in the images provide a
lower bound for the length scale of equilibrium dislocation loops.
We discuss this point below after presenting a quantitative
analysis of the translational order.

To study quantitatively the FL lattice order, we proceed as
follows. First, a perfect lattice is constructed and registered to
the FL positions obtained from an experimental image.  The initial
lattice vectors used to construct the perfect lattice were
obtained from the Fourier transform of the vortex positions.  When
an image contains a dislocation, the continuum approximation is
used to construct the perfect lattice with the dislocation
\cite{[15]}.  Next we minimized the root mean square displacement
between the underlying perfect lattice and the real FL lattice by
varying the position and orientation of the two lattice vectors of
the perfect lattice.  The displacement vector
$\mathbf{u}(\mathbf{r})$ associated with each of the vortices
positioned at $\mathbf{r}$ relative to the perfect lattice was
then computed.  Fig~\ref{fig1} displays a color-representation of
the displacement field for a typical dislocation-free image and an
image containing three dislocations.  In Fig~\ref{fig1}(a), the
average displacement is 0.22 $a_{0}$, where $a_{0}$ is the lattice
constant. Qualitatively, the map consists of several intermixed
domain-like structures, within which the displacement fields are
correlated. These uniformly dispersed domain-like structures of
the displacement field produce sharp Bragg peaks in Fourier space
(see Fig~\ref{fig3}(b) later).  We also believe that
$\mathbf{u}(\mathbf{r})$ provides a quick indication of
nonequilibrium effects.  For example, Fig~\ref{fig1}(b) exhibits
large domains of correlated displacements that are sheared
relative to each other; that is, the blue-green-blue coded
domains. We believe that this larger scale distortion is a
manifestation of a nonequilibrium structure that may arise from
quenched dynamics of FLs during our field-cooling process (see
below).

To compare our data directly with theoretical predictions
\cite{[5]}, we have calculated the displacement correlator,
$B(r)$, and translational correlation function,
$C_{\mathbf{G}}(r)$. $B(r)$ and $C_{\mathbf{G}}(r)$ are defined as
$\langle[\mathbf{u}(\mathbf{r})-\mathbf{u}(\mathbf{0})]^{\mathrm{2}}
\rangle/\mathrm{2}$ and $\langle
e^{i\mathbf{G}[\mathbf{u}(\mathbf{r})-\mathbf{u}(\mathbf{0})]}\rangle$,
respectively, where $\langle\;\rangle$ is the average over thermal
fluctuations and quenched disorder, and $\mathbf{G}$ is one of the
reciprocal lattice vectors. Theoretically \cite{[5]}, we expect
$B(r)$ will show three distinct behaviors as $r$ increases:
$B(r)\sim r$ in the Larkin regime, where $B(r)$ is less than the
square of $\xi$, the in-plane coherence length.  As $r$ increases
further, FLs are in the random manifold regime where $\xi^{2} <
B(r) < a_{0}^{2}$. In this regime $B(r) \sim r^{2\nu}$ with the
roughening exponent ${2\nu}$ ($< 1$).  Finally, at the largest
length scales (the quasiordered regime) where $a_{0}^{2} < B(r)$,
$B(r)\sim \ln r$. Since the in-plane $\xi$ of BSCCO is only $\sim$
20 \AA, the Larkin regime is irrelevant in our experiment (i.e.,
$a_{0}\gg \xi$).  Fig~\ref{fig2}(a) shows the behavior of B(r)
calculated from the data in Fig~\ref{fig1}(a). For $r < 80 a_{0}$,
$B(r)$ can be fit well with a power law, $B(r) \sim r^{2\nu}$,
with $2\nu = 0.44$.  Thus our experiment is probing the random
manifold regime at least up to this scale.  Indeed, $B(r)$ grows
only up to 0.05 $a_{0}^{2}$ at $r = 80 a_{0}$, well below the
expected crossover to the quasiordered regime, i.e.  $B(r) \sim
a_{0}^{2}$. A naive extrapolation to $B(r) = a_{0}^{2}$ suggests
the crossover at $r \sim 10,000 a_{0}$ ($\sim$ 4 mm), which is far
beyond our experimental limit. Samples with such a large clean
area, and direct imaging of $\sim$ $10^{8}$ vortices would be
required to observe the logarithmic roughening of FLs. The
roughening exponent $2\nu$ is found to be independent of the field
(70 - 120 G) and consistent with the estimate $2\nu = 2/5$
obtained by FeigelÕman et al. using a scaling argument \cite{[3]}.
As shown in Fig~\ref{fig2}(b), $C_{\mathbf{G}}(r)$ and
$e^{-G^{2}B(r)/2}$ overlap with each other for $r < L^{*}$, where
the measured $L^{*}$ is $\sim 80 a_{0}$. These results support the
Gaussian approximation, $C_{\mathbf{G}}(r) \approx
e^{-G^{2}B(r)/2}$, which has been simply assumed for the
equilibrium FLs lattice \cite{[5]} within this length scale.  For
$r > L^{*}$, however, B(r) deviates strongly from expected
behavior; that is,$B(r)$ saturates and even decreases as $r$
increases.  In addition, the Gaussian approximation breaks down
for $r > L^{*}$ as evidenced by the difference between
$C_{\mathbf{G}}(r)$ and $e^{-G^{2}B(r)/2}$.  We believe that this
behavior can be attributed to nonequilibrium FL structures at the
larger length scales of our experiment.

To examine this point further, we decompose $B(r)$ into its
longitudinal [$B^{L}(r)$] and transverse [$B^{T}(r)$] parts:
$B(r)=(B^{L}(r)+B^{T}(r))/2$, where
\begin{equation}
B^{L}(r)=\langle\Bigl(\bigl(\mathbf{u}(\mathbf{r})-\mathbf{u}(0)\bigr)
\cdot\frac{\mathbf{r}}{\mathnormal{r}}\Bigr)^{\mathrm{2}}\rangle.
\end{equation}
It is worth noting that in the random manifold regime, the ratio
of $B^{T}(r)$ and $B^{L}(r)$ is predicted to be $2\nu +
1$~\cite{[5]}, and thus an independent estimate of the roughening
exponent.  The average value of this ratio measured from our data
(inset to Fig~\ref{fig3}(a)) is 1.40, which is consistent with the
value of $2\nu$ obtained from $B(r)$.  As shown in
Fig~\ref{fig3}(a), both $B^{L}$ and $B^{T}(r)$ are described well
with the power law behavior up to $r \sim L^{*}$. Beyond this
range, however, the transverse displacement $B^{T}(r)$ first
deviates from power law causing deviations in $B(r)$. Thus, we
infer that shear motion of FL lattice should be responsible for
the abnormal behavior of $B(r)$. Since the shear modulus of FL
lattice is much smaller in magnitude than the compressional
modulus \cite{[1]}, $B^{T}(r)$ is always larger than $B^{L}(r)$,
and the shear motion dominates the relaxation of the FL lattice
during the field cooling process.  As temperature decreases, the
long wavelength component of shear motion is frozen out.  We
believe that the domain-like structures seen in Fig~\ref{fig1} are
a snap shot of these frozen long wavelength shear motions. Note
that the characteristic length scale of these domain like
structures in Fig 1 is again $\sim L^{*}$, which explains the
deviations in $B(r)$ for $r > L^{*}$. Therefore, $L^{*}$ is the
equilibrium length scale within which FLs can relax to the local
equilibrium during our experimental time scale.

This issue can also be addressed through Fourier space analysis.
Fig~\ref{fig3}(b) displays a blow-up of one Bragg peak.  Several
small satellite peaks appear around the relatively sharp main
peak; these satellite peaks indicate a large-scale modulation of
the FL lattice. If the FLs were in equilibrium, only one main peak
should be expected. The corresponding real space distance between
the main and satellite peaks is, again, $\sim L^{*}$.  Hence these
satellite peaks provide another evidence of the frozen-in dynamics
beyond the equilibrium length scale $L^{*}$.  In addition, we have
prepared FL lattices in different ways to address the
nonequilibrium structures.  For example, we cooled the samples in
the absence of a field to 65 K, applied a field 70 G, and then
cooled slowly (0.1 K/min) to 4 K. Significantly, we find a similar
density of dislocations and FL structure compared to the rapid (10
K/s) field-cooled samples.  Since 65 K is far below the melting
temperature \cite{[17]}, this observation suggests that the
nonequilibrium structures originate from the frozen-in dynamics
far below the melting temperature. Although we can probe FLs up to
a length scale of $\sim$ 300 $a_{0}$, there is a much smaller
length scale $L^{*}$ that prohibits direct application of the
theory derived for an equilibrium FLs. Further studies should
address this important issue.

Finally, we consider the origin of dislocations observed in our
experiments, since nonequilibrium vs.  equilibrium nature of
dislocations is critical to the existence of the Bragg glass
phase. We believe that our data, which exhibit the small numbers
of dislocations, in fact, favors nonequilibrium nature of
dislocation in the FL lattice we probed by following reasons.
First, it is found that most dislocations are pinned in between
domain boundaries (see Fig~\ref{fig1}(b) for example). If there
were a dislocation within the domain-like structures where FLs are
locally in equilibrium, the dislocation should be an equilibrium
feature. Second, $L^{*} \ll L_{d} = n_{d}^{-1/2}\sim250 a_{0}$,
where $L_{d}$ and $n_{d}$ are the average distance between
dislocations and the density of dislocations, respectively. If
dislocations were energetically favorable in an equilibrium FL
lattice, large dislocation loops should proliferate beyond the
equilibrium length scale $L^{*}$. In addition, if some
dislocations drift within domains, and are pinned at domain
boundaries, we should have $L_{d} \lesssim L^{*}$. Therefore, our
experiment ($L^{*} \ll L_{d}$) suggests that dislocations are not
equilibrium features in the FL lattice. Together, our data provide
a lower bound for the length scale of equilibrium dislocation loop
in the FL lattice.

In summary, we have obtained large scale dislocation-free images
of the FL lattice in high quality BSCCO superconductors.
Quantitative analyses of the translational order indicate that the
system is in equilibrium for length scales up to $\sim 80 a_{0}$,
and that FLs are in the random manifold regime with a roughening
exponent $2\nu = 0.44$.  We suggest that the very small density of
dislocations observed in our data is an out-of-equilibrium feature
due to the short time scales involved in our field-cooled
experiments.

We thank D. R. Nelson, D. S. Fisher, P. Le Doussal, and T.
Giamarchi for helpful discussion. CML acknowledges support of this
work by the NSF Division of Materials Research.

\begin{figure}
\vskip 7mm \caption{(color) Spatial map of vortex displacements
$\mathbf{u}(\mathbf{r})$ from the perfect lattice positions.
Darker (brighter) regions in the map correspond to smaller
(larger) displacements.  Different colors correspond to vortex
displacements in different directions, as shown in the inserted
color wheel.  The two solid lines, inner and outer circles in the
color wheel correspond to two basis vectors of the lattice,
displacements of 0.5 $a_{0}$ and $a_{0}$ respectively. Samples
were decorated at 70 G. (a) dislocation-free image containing
37003 vortices.  The edge of the image correspond to 106 $\mu$m.
The lower inset shows a part of both real FL image and a perfect
lattice (yellow) with displacement vectors (red). (b) larger scale
image containing three dislocations (highlighted by red dots and
circles) and large scale shearing.  The image contains 78385
vortices in the 160 x 160 $\mu$$m^{2}$ area.} \label{fig1}
\end{figure}

\begin{figure}
\vskip 7mm \caption{(a) Mean-square relative displacement
correlator $B(r)$ (open circles) as a function of distance r
calculated from the image shown in Fig 1(a).  The solid line is a
power law fit : $B(r) \sim r^{0.44}$.(b) Translational correlation
function $C_{\mathbf{G}}(r)$ (dots) calculated from the same
image.  The open circles are a comparison with the Gaussian
approximation : $C_{\mathbf{G}}(r) \approx e^{- G^{2}B(r)/2}$}
\label{fig2}
\end{figure}

\begin{figure}
\vskip 5mm \caption{ (a) Transverse (open circle) and longitudinal
(solid circle) displacement correlators, i.e. $B^{T}(r)$ and
$B^{L}(r)$, as a function of distance calculated from Fig.\
\ref{fig1}(a). The insert shows the ration of the two quantities.
(b) The Fourier space image showing all six first order Bragg
peaks. (c) Detailed of one of the Fourier peaks calculated from
the same image (inverted gray scale). Dark arrows highlight two
satellite peaks.} \label{fig3}
\end{figure}

\end{multicols}
\end{document}